\pdfoutput=1
\documentclass[aps,prb,reprint,
superscriptaddress]{revtex4-2}
\usepackage[utf8]{inputenc}
\usepackage{hyperref}
\usepackage{xcolor}

\usepackage{graphicx}
\graphicspath{{./}}
\usepackage{dcolumn}
\usepackage{bm}
\usepackage{booktabs}
\usepackage{amsmath}
\usepackage{amssymb}

\begin{document}

\title{Approximate label symmetries improve data efficiency}

\author{Scott Y. H. Kim}
\affiliation{Department of Chemistry, Chemical Physics Theory Group, University of Toronto, St. George Campus, Toronto, ON M5R 0A3, Canada}
\author{Mathis Lechaume-Robert}
\affiliation{Institut Courtois, Université de Montréal, Montréal, QC H2V 0B3, Canada}
\affiliation{Acceleration Consortium, University of Toronto, Toronto, ON M5R 0A3, Canada}
\author{O. Anatole von Lilienfeld}
\email{anatole.vonlilienfeld@utoronto.ca}
\affiliation{Department of Chemistry, Chemical Physics Theory Group, University of Toronto, St. George Campus, Toronto, ON M5R 0A3, Canada}
\affiliation{Acceleration Consortium, University of Toronto, Toronto, ON M5R 0A3, Canada}
\affiliation{Department of Materials Science and Engineering, University of Toronto, St. George Campus, Toronto, ON M5R 0A3, Canada}
\affiliation{Department of Physics, University of Toronto, St. George Campus, Toronto, ON M5R 0A3, Canada}
\affiliation{Vector Institute for Artificial Intelligence, Toronto, ON M5S 1M1, Canada}

\date{\today}

\begin{abstract}
  Enforcing feature symmetries in machine learning (ML) models is a common strategy to mitigate data scarcity.
  Confirming expectations from statistical learning theory, we show that  exact, as well as approximate, {\em label} symmetries can also improve data efficiency.
  We illustrate the idea for the $s, p, d$ orbital densities of the electron in the hydrogen atom, for the three vibrational normal modes of the water molecule, and for its full 3D potential energy hypersurface.
  Resulting ML models of electron density and potential energies exhibit superior learning curves, demonstrating improved generalization efficiency.
  We observe that learning curves similarly improve even when label symmetries are not exact --- up to the convergence floors set by the degree to which the symmetry is approximate.
  Further improvements are obtained for approximate label symmetries in the molecular potential energy surface, using a Hessian-based correction that suppresses the leading order term in the error.
  \end{abstract}

  \maketitle

  \section{Introduction}
  \begin{figure}
    \centering
    \includegraphics[width=\columnwidth]{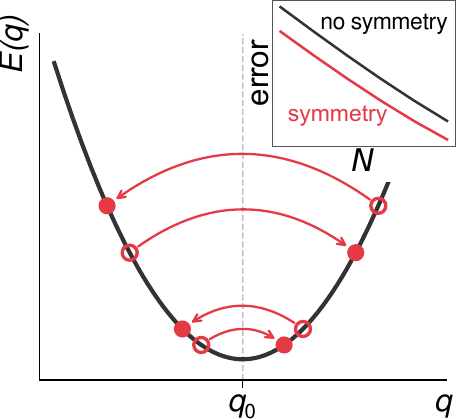}
    \caption{Schematics of label-symmetry enforcement for data augmentation resulting in improved ML model training data efficiency.
    For any even target function $E(q)$, each input $q_i$ (hollow markers) is related to $-q_i$ (filled markers) by the reflection group and approximately carries the same label.
    Reflection of features about $q_0$ yields symmetry-related points which can be used to augment the training dataset without incurring additional acquisition cost.
    ML models trained with all empty and filled points exhibit a constant-factor improvement in data efficiency without a change in learning rate (see inset).}
    \label{fig:intro}
    \end{figure}

  Machine learning (ML) models trained on quantum-mechanical (QM) reference data across chemical compound space deliver  property predictions at a fraction of first-principles cost with generalization errors that scale with training set size \cite{CM,QMLessayAnatole,faber2017prediction}.
    Despite the success of query-aware similarity ML methods~\cite{lemm2023improved}  or large foundation models~\cite{foundation},
  their fundamentally interpolative nature reduces their predictive power
  for out-of-distribution queries~\cite{li2023critical}.
  Because high-fidelity QM labels are expensive at the scale required for
  reliable explorations of chemical spaces, inductive biases that increase training data  efficiency remain desirable.

  Symmetry is a way to improve data efficiency.
  For example, invariances and equivariances of scalar and vectorial properties such as energies and forces under $SE(3)$ and permutations of identical nuclei have
  been exploited at the level of internal coordinate representations~\cite{Neuralnetworks_BehlerParrinello2007, Neuralnetworks_Behler2011,
  PIP, CM, BartokGabor_Descriptors2013, GDML, ANI1, FCHL, drautz2019atomic} or  through equivariant
  message-passing architectures~\cite{SchNet, NewtonNet, EGNN, PaiNN, NequIP, MACE}.
  A closely related
  strategy used in computer vision and natural language
  processing is based on augmenting training data with symmetry-related configurations
  ~\cite{AlexNET, shorten2019survey, dao2019kernel, chen2020group}.
  Empirically, symmetry enforcing representations result in a constant factor improvement in data efficiency, while architectural enforcement of $SE(3)$ equivariance has
  been observed to accelerate the learning rate itself~\cite{NequIP,Ngo2025}.

  $SE(3)$ invariance is a universal symmetry of the molecular Hamiltonian and holds for every observable automatically.
  Label symmetries, as we use the term here, are additional degeneracies of the specific label function that do not necessarily follow from those universal symmetries.
  They relate configurations that can not be connected by global translation, rotation, or
  index permutation, but nonetheless are approximately degenerate.
  Not all molecular properties carry such symmetries, and identifying one typically requires prior knowledge of the qualitative functional form of the target itself.
  To the best of our knowledge, these symmetries have not yet been exploited in the context of atomistic machine learning.

  In practice, exact label symmetries are rare.
  This differs from approximate symmetrization currently used in the
literature~\cite{tahmasebi2025approximate,langer2024probing,domina2025unconstrained}, where  the label is exactly symmetric and the model enforces it only approximately.
When the label itself is  not exactly symmetric, even a perfect machine learning model would still reproduce the noise due to symmetry breaking.

  To illustrate our point we have selected two different problems that are fundamental for physical chemistry:
  For exact label symmetries, we first consider the electron orbital densities in the hydrogen atom, which admit discrete reflection and continuous rotation label symmetries.
  For approximate label symmetries, we focus on reflection symmetries of the molecular potential energy well about the equilibrium geometry of the water molecule.
  \section{Theory}

  \subsection{Learning Curves}
  Learning curves plot prediction error $\epsilon$ against training set size $N$ and
  are the standard data-efficiency benchmark for ML models in quantum chemistry and
  materials science~\cite{CM,faber2017prediction,FCHL,Huang2021chemrev}.
  Multiple theoretical frameworks have consistently predicted an inverse power-law decay
  of $\epsilon$ with $N$~\cite{blumer1989VC,ehrenfeucht1989VC,baum1988VCpredict,
  haussler1991VCpredict,levin1990statistical,seung1992statistical,
  rissanen1986stochastic,amari1992four,amari1993LCSM,Cortes1993,Muller1996},
  confirmed empirically in many settings using neural networks~\cite{Hestness2017}
  and kernel methods~\cite{Spigler2020,Li2023KRR}.

  We therefore model the learning curve by
  \begin{align}
    \epsilon(N) - \epsilon_\infty \approx A N^{-\beta},
    \label{eq:lc_powerlaw}
  \end{align}
  where $\epsilon_\infty = \lim_{N\to\infty}\epsilon(N)$ is the asymptotic error floor,
  $A$ is a prefactor, and $\beta$ is the power-law exponent.
  For deterministic, exactly labeled data $\epsilon_\infty = 0$;
  inconsistent or noisy labels raise it to a positive value that additional data cannot
  reduce~\cite{amari1993floor,rosenfeld2019constructive}.
  In the approximate-symmetry setting studied below, augmenting with mirror points assigns
  deterministically incorrect labels, producing a nonzero $\epsilon_\infty$ even without
  stochastic noise.

  \subsection{Data Efficiency Improvements from Exact Symmetries}
  When symmetries are exact, as in the electron orbital densities in the hydrogen atom, improved data efficiency persists at all training set sizes.

  Symmetry acts on $A$ and $\beta$ through different mechanisms depending on whether
  it is continuous or discrete.

  Continuous symmetries reduce the effective dimensionality of the learning problem.
  Since learning rates for smooth functions improve with lower
  dimensionality~\cite{tahmasebi2023exact}, continuous symmetries increase $\beta$ rather than
  only the prefactor $A$; higher levels of $SE(3)$ equivariance yield systematically
  steeper learning curves empirically~\cite{NequIP,Ngo2025}.

  By contrast, discrete symmetries effectively multiply the information content of each
  training point by a factor of $b$, the number of points related by the symmetry,
  provided no two training points are already related by symmetry.
  With training set size $N$, and in the power-law regime where $\epsilon_\infty$ is negligible,
  \begin{align}
    \ln\epsilon(bN) = \ln A - \beta\ln b - \beta\ln N,
  \end{align}
  and comparison to the unsymmetrized model gives
  \begin{align}
    \ln\left(\frac{\epsilon(bN)}{\epsilon(N)}\right) = -\beta\ln b,
    \label{eq:lc_discrete_shift}
  \end{align}
  a constant independent of $N$, so discrete symmetry acts as a pure horizontal shift
  by $\ln b$ on a log-log plot.
  Equation~\eqref{eq:lc_discrete_shift} holds only in the pre-floor regime; $\epsilon_\infty$
  for each augmentation scheme is discussed in the following section using a Taylor expansion
  of the approximate mirror labels.

  \subsection{Data Efficiency Improvements from Approximate Symmetries}
  \label{sec:parity}
  In the potential energy well about the equilibrium geometry of the water molecule, the discrete reflection $\mathbf{q}\mapsto -\mathbf{q}$, where $\mathbf{q}$ is the vector of mass-weighted normal-mode displacements from equilibrium, is an approximate label symmetry due to the anharmonicity of the vibrational normal modes. In this section, we argue that the approximate nature of the reflection symmetry introduces an error floor that bounds improvements in data efficiency.

  \subsubsection{Label errors from augmentation}
  In one dimension, the electronic energy $E$ in terms of $q$ is
  \begin{align}
    E(q) &= E(0) + \frac{1}{2} q^2 \partial^2_q E_0 + \frac{1}{6} q^3 \partial^3_q E_0 + \frac{1}{24} q^4 \partial^4_q E_0 + \mathcal{O}(q^5),
    \label{eq:mode_expand}
  \end{align}
  where $\partial^n_q E_0 = \left.\left(\frac{\partial^n E}{\partial q^n}\right)\right|_{q=0}$. In a standard frequency analysis,
  $\partial^2_q E_0$ is available as the force constant of the vibrational mode. Moving forward, we will denote
  $E_{\text{HO}}(q) = E(0) + \frac{1}{2}q^2\partial^2_q E_0$.
  The energy of the reflected point $-q$ is
  \begin{align}
    E(-q) &= E_{\text{HO}}(q) - \frac{1}{6} q^3 \partial^3_q E_0 + \frac{1}{24} q^4 \partial^4_q E_0 + \mathcal{O}(q^5),
    \label{eq:mode_reflect}
  \end{align}
  differing only in sign in the cubic term.

  With the knowledge that the electronic energy is approximately $\mathbb{Z}_2$ symmetric about the equilibrium geometry,
  each training point $(q,E(q))$ is paired with a mirror point at $-q$ carrying the same label
  \begin{align}
    \tilde{E}^{\text{sym}}(-q) &= E(q).
    \label{eq:sym_mirror_label}
  \end{align}
  The error of this approximately symmetric label is
  \begin{align}
    \tilde{E}^{\text{sym}}(-q) - E(-q) &= \frac{1}{3} q^3 \partial^3_q E_0+ \mathcal{O}(q^5),
    \label{eq:symm_label_error}
  \end{align}
  for which all even order terms have vanished, in analogy to the recently developed anti-symmetric alchemical perturbation approach~\cite{anatole2025diastereomers}.
  The label function generated by this augmentation procedure is called $\mathrm{Aug}_2$ in the following.
  It is even by construction, and has a leading third order error term which must introduce an irreducible generalization error floor, roughly corresponding to the cubic term in the expansion.

  A correction for the $\mathrm{Aug}_2$ estimate of $E(-q)$ can be obtained by first adding Eq.~\ref{eq:mode_expand} to Eq.~\ref{eq:mode_reflect}, and rearranging,
  \begin{align}
    E(-q) &= 2E_{\text{HO}}(q) - E(q) + \frac{1}{12} q^4 \partial^4_q E_0 + \mathcal{O}(q^6).
    \label{eq:mirror_expand}
  \end{align}
  Then, setting
  \begin{align}
    \tilde{E}^{\text{corr}}(-q) &= 2E_{\text{HO}}(q) - E(q),
    \label{eq:taylor_mirror_label}
  \end{align}
  we obtain the error of the corrected label up to fourth order:
  \begin{align}
    \tilde{E}^{\text{corr}}(-q) - E(-q) &= -\frac{1}{12} q^4 \partial^4_q E_0 + \mathcal{O}(q^6),
    \label{eq:taylor_label_error}
  \end{align}
  Equation~\eqref{eq:taylor_mirror_label} has the structure of a single position-Verlet
  step~\cite{Verlet1967}, using $E$ and $q$ as coordinates instead of position and time.
  This augmentation scheme is a special case where the first-order term is already $0$ by virtue of using a stationary point as reference.
  The corresponding corrected label is dubbed $\mathrm{Aug}_3$ in the following.

 \subsubsection{Generalization to $d$ dimensions}
  The data augmentation construction generalizes to $d$ normal-mode coordinates as follows.
  Writing $\mathbf{q}=(q_{1},\ldots,q_{d})$, the harmonic reference
  \begin{align}
    E_{\mathrm{HO}}(\mathbf{q})
      = E(0) + \frac{1}{2}\sum_j k_j\, q_j^2
    \label{eq:ho_nd}
  \end{align}
  is invariant under $\mathbf{q}\mapsto-\mathbf{q}$, where $k_j$ is the force constant.
  Extending Eq.~\eqref{eq:mode_expand}, odd-order cross-terms change sign under reflection  while even-order terms do not, giving
  \begin{align}
    E(-\mathbf{q})
      &= E_{\mathrm{HO}}(\mathbf{q})
         - \frac{1}{6}\sum_{klm} q_k q_l q_m \left.\frac{\partial^3 E}{\partial q_k \partial q_l \partial q_m}\right|_0 \nonumber\\
      &\quad{}
         + \frac{1}{24}\sum_{klmn} q_k q_l q_m q_n \left.\frac{\partial^4 E}{\partial q_k \partial q_l \partial q_m \partial q_n}\right|_0
         + \mathcal{O}(q^5).
    \label{eq:mode_reflect_nd}
  \end{align}
  The symmetric mirror label $\tilde{E}^{\text{sym}}(-\mathbf{q})=E(\mathbf{q})$
  has error
  \begin{align}
    \tilde{E}^{\text{sym}}(-\mathbf{q}) - E(-\mathbf{q})
      &= \frac{1}{3}\sum_{klm} q_{k}\,q_{l}\,q_{m} \left.\frac{\partial^3 E}{\partial q_k \partial q_l \partial q_m}\right|_0 \nonumber\\
      + \mathcal{O}(q^5),
    \label{eq:symm_label_error_nd}
  \end{align}
  the multidimensional analogue of Eq.~\eqref{eq:symm_label_error}.

  Since $E(\mathbf{q})+E(-\mathbf{q})$ cancels all odd-order terms,
  the corrected mirror label
  $\tilde{E}^{\text{corr}}(-\mathbf{q})=2E_{\mathrm{HO}}(\mathbf{q})-E(\mathbf{q})$ gives
  \begin{align}
    \tilde{E}^{\text{corr}}(-\mathbf{q}) - E(-\mathbf{q})
      &= -\frac{1}{12}\sum_{klmn} q_{k}\,q_{l}\,q_{m}\,q_{n} \nonumber\\
      &\qquad{}\times \left.\frac{\partial^4 E}{\partial q_k \partial q_l \partial q_m \partial q_n}\right|_0
      \nonumber\\
      &\quad{}+ \mathcal{O}(q^6),
    \label{eq:taylor_label_error_nd}
  \end{align}
  the analogue of Eq.~\eqref{eq:taylor_label_error}.
  Both errors retain the same cubic/quartic structure as in one dimension
  but are generally larger in magnitude due to the inclusion of cross-mode coupling.

  These label errors determine the irreducible convergence floors of generalization errors of machine learning models trained on augmented data using $\mathrm{Aug}_2$ and $\mathrm{Aug}_3$.

  \subsubsection{Convergence floors}
  \label{sec:floors}
  In the $N\to\infty$ limit, every point $q$ is labeled with the
  true label $E(q)$ as well as an augmented label from the mirror of the
  point at $-q$.
  For symmetric augmentation $\mathrm{Aug}_2$, the augmented label for $d=1$ is  $\tilde{E}^{\text{sym}}(q)=E(-q)$.
  The squared-loss minimizer over both labels at $q$ is their average
  \begin{align}
    \hat{f}^{\text{sym}}(q)
      = \frac{E(q)+E(-q)}{2},
    \label{eq:sym_asymp}
  \end{align}
  the definition of the even component of the energy function.
  The irreducible floor in symmetric augmentation must then be the
  odd component of $E(q)$, which to leading order in the Taylor expansion is
  \begin{align}
    \epsilon_\infty^{\mathrm{sym}}
      \approx \tfrac{1}{6}\bigl\langle|q^3|\bigr\rangle|\partial^3_q E_0|,
    \label{eq:sym_floor}
  \end{align}
  where angle brackets denote the average over the test distribution.
  For the corrected scheme ($\mathrm{Aug}_3$), the mirror of the point at $-q$
  contributes the corrected label $\tilde{E}^{\mathrm{corr}}(q) = 2E_{\mathrm{HO}}(q)-E(-q)$.
  The average of this label and the true label $E(q)$ is
  \begin{align}
    \hat{f}^{\mathrm{corr}}(q)
      &= E_{\mathrm{HO}}(q) + \frac{E(q)-E(-q)}{2},
    \label{eq:corr_asymp}
  \end{align}
  which contains the harmonic baseline and the odd component of the
  anharmonic residual.
  The irreducible floor is determined by the even component of the
  residual, which to leading order is
  \begin{align}
    \epsilon_\infty^{\mathrm{corr}}
      \approx \tfrac{1}{24}\bigl\langle|q^4|\bigr\rangle|\partial^4_q E_0|,
    \label{eq:taylor_floor}
  \end{align}
  a floor that is generically smaller than Eq.~\eqref{eq:sym_floor}.

  \section{Methods}

  \subsection{Symmetry enforcement}
  \label{sec:symm_methods}
  A label symmetry can be enforced in one of three ways. The first is data
  augmentation, which adds synthetic data related by symmetry to the original training set.
  The second is input transformation, which replaces $x$ by a representation $M(x)$ that is invariant to the symmetry, realized either as a
  canonicalization to a fundamental domain or as an invariant descriptor
  coordinate~\cite{Neuralnetworks_BehlerParrinello2007, Neuralnetworks_Behler2011,
  PIP, CM, BartokGabor_Descriptors2013, FCHL,drautz2019atomic}.

  When the symmetry is exact, both strategies are equivalent and produce
  identical learning curves, which we confirm on a model even target in the
  Supplemental Material (Fig.~S2). We therefore adopt whichever is most convenient
  for each system. For the hydrogen orbital densities the symmetry is exact and a
  fundamental-domain coordinate is available in closed form, so we use input
  transformation. For the approximate reflection symmetry of the water potential
  energy surface the mirror label is not exact, so we use augmentation, which lets
  the reflected points carry the $\mathrm{Aug}_2$ or $\mathrm{Aug}_3$ labels
  developed in Sec.~\ref{sec:parity}.

  \subsection{Hyperparameters}

  All experiments were performed using kernel ridge regression (KRR)~\cite{muller2001introduction}. Hyperparameters were selected by 4-fold cross-validation
  on a logarithmically spaced grid. The regularization was
  $\lambda\in[10^{-14},10^{-6}]$ with 5 values for the hydrogen-orbital and 1D
  water experiments, and $\lambda\in[10^{-10},10^{-3}]$ with 8 values for the 3D
  water experiments.
  For the $s$-orbital experiments, the Laplacian length-scale grid was
  $\sigma\in[10^{-2},10^{2}]$ with 10 values.
  For the $p_z$ and $d_{xz}$ experiments, $\sigma\in[10^{-1.5},10^{0.5}]$ with
  10 values.
  For the 1D water normal-mode scans, the Laplacian grid was
  $\sigma\in[10^{-2.5},10^{0}]$ with 10 values; for the 3D water experiments, the
  Gaussian grid was $\sigma\in[10^{-2},10^{0.6}]$ with 13 values.

  \subsection{Water datasets}

  Vibrational normal modes and force constants were obtained from a frequency analysis.
  For the 1D scans, 1000 geometries along each normal mode were generated by drawing displacements uniformly from $[-q_{\max,j},\,q_{\max,j}]$, where $q_{\max,j} = \sqrt{k_BT/k_j}$ for $T = 1500$~K.
  For the 3D dataset, 10000 structures were generated by simultaneous displacement along all three
  normal-mode coordinates $\mathbf{Q}=(q_1,q_2,q_3)$, with each $q_j$ drawn independently
  from a Gaussian with variance $\sigma_j^2 = k_BT/k_j$ at $T=300$~K.
  Mirror points were constructed by inversion to $-\mathbf{Q}$, with labels assigned
  by $\mathrm{Aug}_2$ or $\mathrm{Aug}_3$.
  Each learning-curve point is the geometric-mean test-set MAE over $100$ random
  training selections, and the shaded confidence bands are $95\%$ bootstrap
  intervals over those selections.
  The odd and even anharmonic components quoted for the 3D floors are the
  degree-3 and degree-4 parts of a least-squares quartic fit of the reference
  residual $\Delta E = E - E_\mathrm{HO}$ over the $1000$-point test set
  (root-mean-square error $0.02$~meV); their mean absolute values give the
  $\mathrm{Aug}_2$ and $\mathrm{Aug}_3$ floors.

  \subsection{Anharmonic vibrational frequencies}
  \label{sec:vpt2}

  Anharmonic fundamentals were obtained from second-order vibrational perturbation
  theory (VPT2)~\cite{Nielsen1951,Mills1972,Barone2005VPT2}. The potential is expanded
  about equilibrium in the dimensionless normal coordinates $q_i$,
  \begin{equation}
    V = \tfrac12\sum_i \omega_i q_i^2
      + \tfrac16\sum_{ijk}\phi_{ijk}\,q_iq_jq_k
      + \tfrac1{24}\sum_{ijkl}\phi_{ijkl}\,q_iq_jq_kq_l ,
    \label{eq:qff}
  \end{equation}
  with harmonic wavenumbers $\omega_i$ and cubic and quartic force constants
  $\phi_{ijk}$ and $\phi_{ijkl}$. In VPT2, the fundamental frequencies are
  determined by the complete cubic force field and the semi-diagonal quartics
  $\phi_{iiii}$ and $\phi_{iijj}$, together with Coriolis and resonance
  corrections~\cite{Mills1972,Barone2005VPT2}. We use the generalized VPT2
  treatment implemented in \textsc{pyVPT2}, including deperturbation and
  variational treatment of Fermi resonances~\cite{Barone2005VPT2,Bloino2012}.

  Calculations used \textsc{pyVPT2}~\cite{pyVPT2}, which assembles the force field of
  Eq.~\eqref{eq:qff} by finite differences along the normal coordinates and performs the
  GVPT2 processing; the harmonic frequencies and modes were taken from the
  $\omega$B97X-D3/def2-TZVPP Hessian. For the learned spectra, the Gaussian KRR
  residual was converted to a local quartic force field by analytic differentiation
  at equilibrium. Taylor coefficients through fourth order were retained, while
  constant, linear, and quadratic residual terms were set to zero so that the
  equilibrium geometry and harmonic frequencies remained fixed by the exact harmonic
  reference. The surface supplied to \textsc{pyVPT2} was therefore
  $E_\mathrm{HO}(\mathbf{q}) + P_{\Delta E}^{(4)}(\mathbf{q})$. Because this
  surface is a quartic polynomial, its finite-difference force field reproduces
  the analytic force constants exactly. Energies were passed through a custom
  provider in place of the quantum-chemistry backend, so no single points are
  evaluated during VPT2 while its finite-difference and resonance machinery run
  unchanged.

  The DFT/VPT2 reference and the cubic/quartic decomposition in Table~\ref{tab:ir}
  were obtained by applying the same quartic projection to the DFT residual
  $E_\mathrm{DFT}-E_\mathrm{HO}$ and rerunning \textsc{pyVPT2} with the cubic and
  full cubic-plus-quartic sectors. The harmonic baseline in the spectra is the
  corresponding zero-residual \textsc{pyVPT2} result, which includes the
  vibration-rotation/Coriolis terms present in the GVPT2 treatment. Dipole moments were computed on the displaced
  three-dimensional water geometries, and the derivatives
  $\partial\boldsymbol\mu/\partial q_i$ were obtained by a least-squares linear fit
  of $\boldsymbol\mu(\mathbf{q})$. Infrared intensities were then evaluated in the
  double-harmonic approximation, and stick spectra were broadened with Lorentzians
  of full width at half maximum $\gamma=18\,\mathrm{cm}^{-1}$.

  \subsection{Software}

  All energy and Hessian calculations of water were
  performed using density functional approximations as implemented in the PySCF 2.8.0 package\cite{pyscf_article} at the $\omega$B97X-D3/def2-TZVPP level of theory\cite{chai2008long,grimme2010consistent,Weigend2006}.
  All machine learning models were implemented in Python using the QML2 machine learning
  package\cite{QML2}, and anharmonic vibrational analyses with \textsc{pyVPT2}~\cite{pyVPT2}.

  \section{Results and Discussion}
  The results are organized in two parts, covering first exact label symmetries on hydrogen atomic orbital densities and then approximate label symmetries on the water potential energy surface.
  \subsection{Exact symmetries}
  We use hydrogen atomic orbital densities to demonstrate the effect of continuous and discrete label symmetries, enforced throughout by input transformation.
  Training points are drawn uniformly in the sphere volume of appropriate radius, and density values are normalized to $[0,1]$ by dividing by the maximum.

  \subsubsection{Continuous symmetries}

  \begin{figure*}
    \centering
    \includegraphics[width=\linewidth]{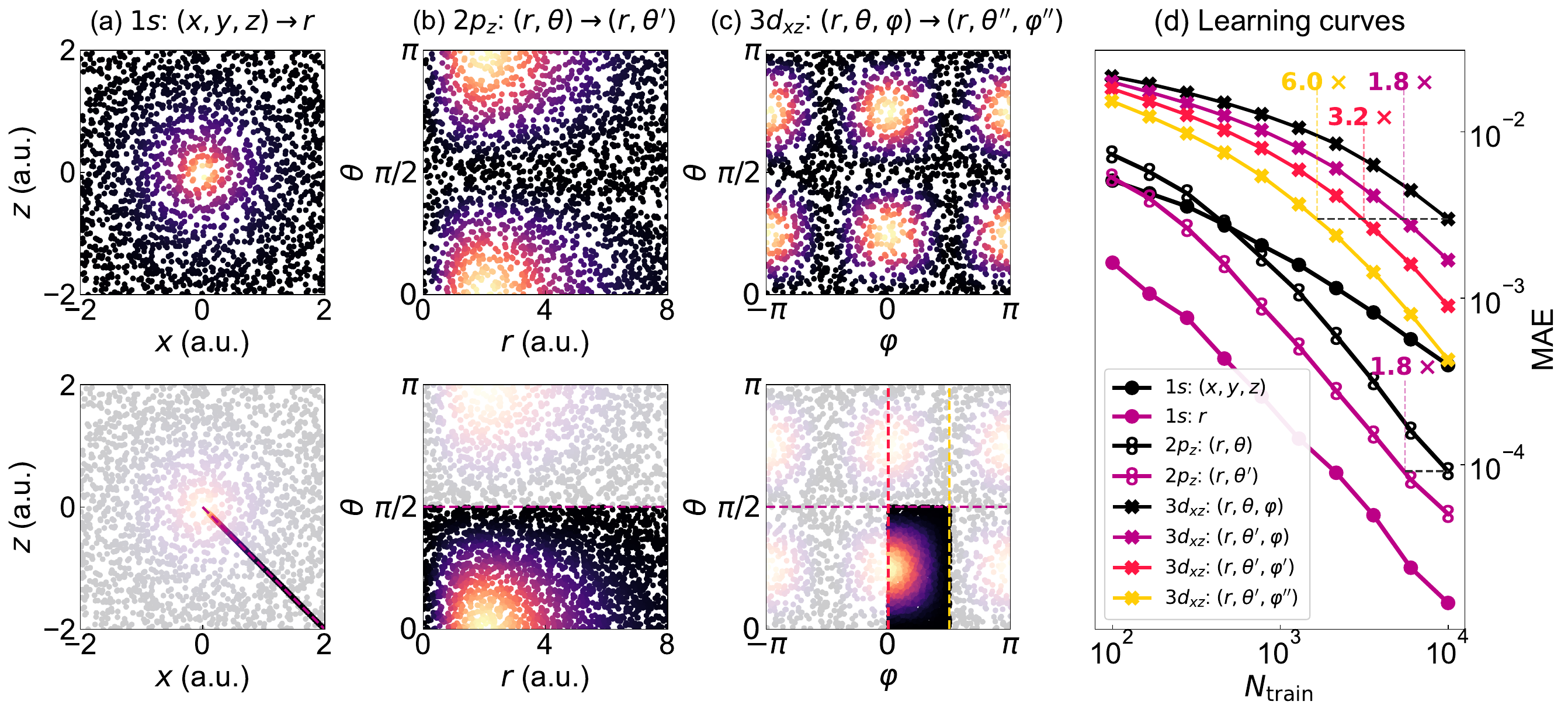}
    \caption{Hydrogen $s$, $p$, and $d$-orbital densities with (top), and without symmetry (bottom).
    Dashed lines in (a), (b), and (c) indicate symmetry planes; faded points mark the redundant domain.
    (a)~$1s$ density $xz$ cross-section at $y=0$ (top) and as a function of $r$ (bottom),    illustrating the continuous reduction from three Cartesian coordinates to one radial coordinate  by rotational symmetry.
    (b)~$2p_z$ density before (top) and after (bottom) enforcing the reflection $\theta\mapsto \theta'$.
    (c) $(\theta,\varphi)$   cross-section of the $3d_{xz}$ density at $r=0$ before (top) and after    (bottom) applying all three reflections.
    (d)~Learning curves for the $1s$, $2p_z$ and $3d_{xz}$
    densities with and without label symmetry enforcement.
    Dashed annotations indicate the data-efficiency gain relative to the default
    models at $N_\mathrm{train}=10\,000$ for the $2p_z$ and $3d_{xz}$ densities.
    }
    \label{fig:hatom}
  \end{figure*}

  Our first demonstration uses the hydrogen $s$-orbital densities
  $\rho_n(r)\propto R_{n0}(r)^2$, where $R_{n0}$ is the radial wavefunction
  with principal quantum number $n$.
  The label symmetry of the target is $O(3)$ invariance, or invariance under
  all rotations of the sphere. We enforce this symmetry by performing
  the regression on the rotation-invariant coordinate $r=\sqrt{x^2+y^2+z^2}$.
  For $n=1,2,3$, we compare KRR models trained on normalized Cartesian
  coordinates $(x,y,z)$ to a symmetry-adapted model trained on
  $r$ alone.

  Learning curves for the $1s$ orbital are shown in Fig.~\ref{fig:hatom}(d); results for the $2s$ and $3s$ orbitals are qualitatively similar and are presented in the Supplemental Material.
  The dominant effect is a change in learning rate. Restricting the fit to the
  asymptotic regime $N_\mathrm{train}\gtrsim 1000$, where the curves approximately follow power-laws, the default models have slopes of $-0.68$, $-0.32$, and $-0.45$ versus
  $-1.15$, $-0.72$, and $-0.58$ in the radially augmented models for $n=1,2,3$, respectively.
  Using $r$ as input (mimicking the effect of spherical data augmentation) yields a steeper learning curve for every orbital, with the  slope steepening by factors of $1.7$, $2.3$, and $1.3$, respectively, despite  differences in functional form and nodal structure.
  The change in slope can be attributed to the dimensionality reduction rather than any feature of the specific orbitals, qualitatively matching the prediction of the effect of continuous symmetries in Ref.~\cite{tahmasebi2023exact}.

  The same dimensionality-reduction mechanism underlies the slope improvements seen in $SE(3)$-equivariant molecular architectures~\cite{NequIP}.
  The distinction is that the $O(3)$ symmetry exploited here is a property of the orbital-density target itself, rather than of the physical space it occupies.

  \subsubsection{Discrete symmetries}
  We next consider discrete point group symmetries in the densities of the $2p_z$
  and $3d_{xz}$ hydrogen orbitals.

  The $2p_z$ density has continuous rotational symmetry about the $z$-axis and is therefore independent of the
  azimuthal angle $\varphi$; the symmetry motivates coordinates $(r, \theta)$ from the outset.
  The remaining discrete symmetry is the reflection through the $xy$ plane.
  Enforcing it replaces $\theta$ with $\theta'=\min(\theta,\pi-\theta)$,
  reducing the $\theta$ domain from $[0,\pi]$ to $[0,\pi/2]$ (Fig.~\ref{fig:hatom}(b)).
  Because spherical coordinates separate, the fold in $\theta$ is independent of $r$.

  In the $3d_{xz}$ density, any point is part of an eight-point set related by three
  reflections across the $xy$, $xz$, and $yz$ planes.
  We enforce these symmetries by defining coordinates $\theta'=\min(\theta,\pi-\theta)$,
  $\varphi'=|\varphi|$, and $\varphi''=\min(|\varphi|,\pi-|\varphi|)$, which sequentially reduce the domains of $\theta$ and $\varphi$ (Fig.~\ref{fig:hatom}(c)).

  The $2p_z$ learning curves (Fig.~\ref{fig:hatom}(d)) exhibit two nearly
  parallel lines in log-log space, with fitted power-law slopes (over $N_\mathrm{train}\geq 1000$) of $-1.22$ for the
  $(r,\theta)$ model and $-1.14$ for the $(r,\theta')$ model.
  The data efficiency is $1.8\times$ higher at
  $N=10\,000$, close to the twofold reduction in the domain caused by
  reflection. For $3d_{xz}$, the models
  are not yet in the fully asymptotic regime across the range studied.
  However, the cumulative data efficiency gain at $N=10\,000$ is $1.8\times$, $3.2\times$, and
  $6.0\times$, corresponding to a per-reflection gain between $1.73$--$1.90$.
  The near-constant gain per reflection suggests that they act nearly
  independently, meaning the total improvement compounds multiplicatively with the
  number of symmetries.
  Whether such independence holds in general depends on the coupling structure of the
  target.
  The hydrogen orbital densities factorize by construction, so each reflection acts on a
  fully decoupled coordinate.
  For a target that does not separate in the folded coordinates, the total gain is expected to fall short of a full factor of 8.

  These results are consistent with the fact that the symmetry is exact, each discrete fold contributes approximately a factor of two to the effective training set size.
  Our second application dealing with the potential energy surface of the molecule examines in the following section  how this picture changes when the symmetry holds only approximately.

  \subsection{Approximate symmetries}
  Here we investigate the impact of approximate symmetries on ML model performance on the PES of water using data augmentation.
  We first study the three individual vibrational normal modes in one dimension, and then across the full three-dimensional PES.

  \subsubsection{1D normal-mode scans}
  \begin{figure*}
    \centering
    \includegraphics[width=\linewidth]{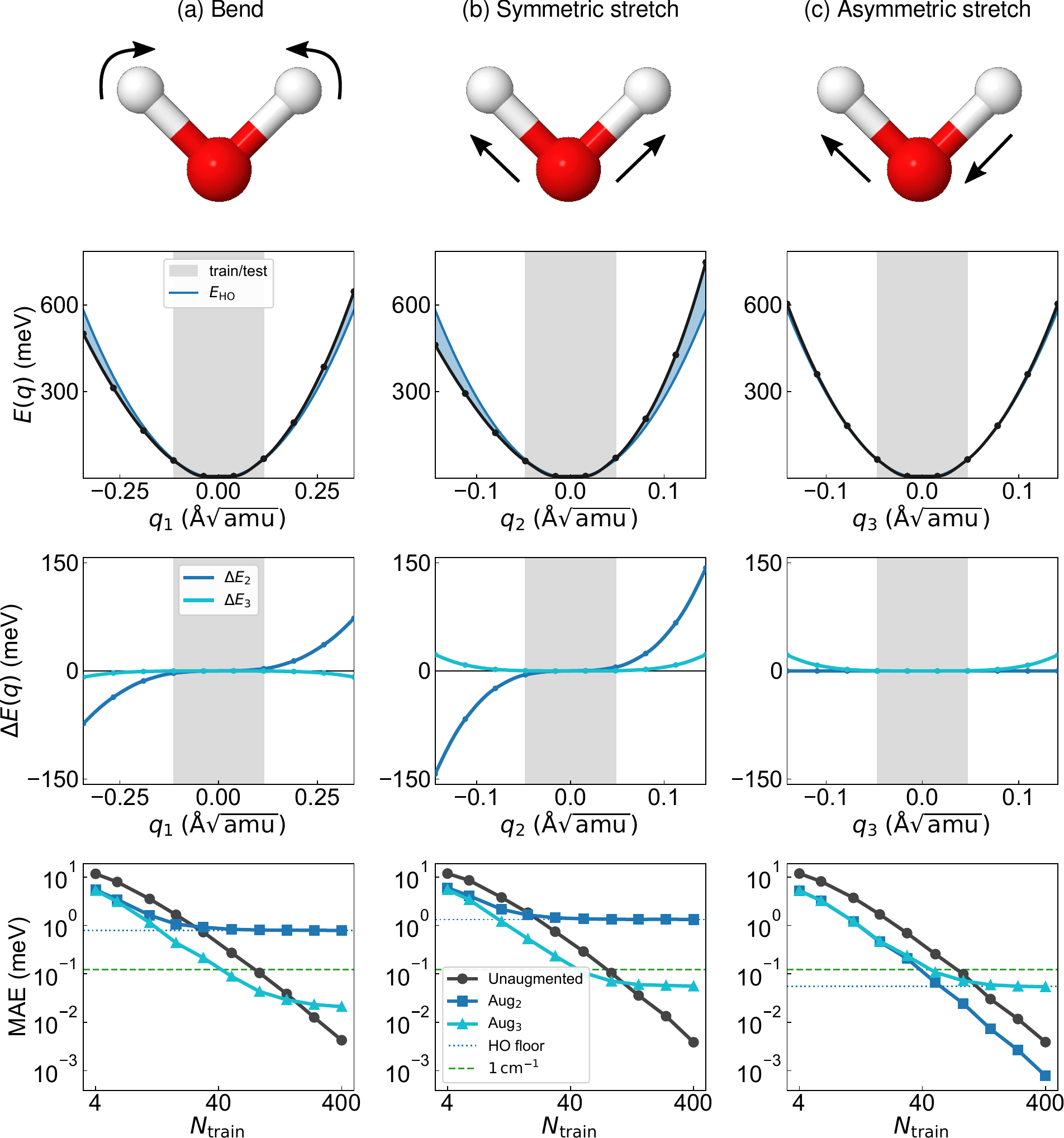}
    \caption{
      Total potential energy learning changes for distortions along
      the (a) bend, (b) symmetric stretch, and (c) asymmetric stretch normal modes of water ($^1$H$_2^{16}$O).
      \emph{Top row:} Schematic of each normal-mode displacement.
      \emph{Second row:} Potential energy $E(q)$ (black) and harmonic reference
      $E_\mathrm{HO}$ (blue dashed) along the mass-weighted normal coordinate $q_i$.
      The shaded blue region denotes the anharmonicity of $E(q)$, while the shaded grey region denotes the region
      sampled during training and testing of ML models.
      \emph{Third row:} the parity components of the anharmonic residual
      $\Delta E = E - E_\mathrm{HO}$---the odd part $\Delta E_2$ (leading cubic) and the
      even part $\Delta E_3$ (leading quartic)---whose test-averaged magnitudes are the
      $\mathrm{Aug}_2$ and $\mathrm{Aug}_3$ floors [Eqs.~\eqref{eq:sym_floor},~\eqref{eq:taylor_floor}].
      \emph{Bottom row:} MAE versus $N_\mathrm{train}$ for unaugmented training,
      symmetric augmentation $\mathrm{Aug}_2$ [Eq.~\eqref{eq:sym_mirror_label}], and corrected augmentation $\mathrm{Aug}_3$ [Eq.~\eqref{eq:taylor_mirror_label}].
      The dotted line marks the harmonic-oscillator anharmonicity floor      $\epsilon_\infty^\mathrm{HO}$, and the dashed line marks an energy of $1\,\mathrm{cm}^{-1}$ ($0.124$~meV).
    }
    \label{fig:water1d}
  \end{figure*}

  We apply label symmetry augmentation to the three vibrational normal modes of water (see Fig.~\ref{fig:water1d}), whose potential energy $E(q)$ is approximately symmetric under $q\mapsto -q$ for the bending and the symmetric stretch mode.
  For the asymmetric stretch mode, the label symmetry is exact.
    Note that this exact symmetry has first been enforced by permutationally invariant polynomials (PIPs)~\cite{PIP}, and label symmetry data augmentation represents  an alternative approach.
    For the two modes with approximate label symmetry, however, there is no alternative to our method.
  Each training point $(q_i, E(q_i))$ is augmented with a mirror point at $-q_i$, with labels assigned by either $\mathrm{Aug}_2$ or $\mathrm{Aug}_3$.

  The anharmonic residuals $\Delta E = E - E_\mathrm{HO}$ of the scans, decomposed in Fig.~\ref{fig:water1d} into odd parts $\Delta E_2$ and even parts $\Delta E_3$, reproduce the qualitative ordering found in empirical and ab initio water force-field studies~\cite{HoyMillsStrey1972,CsaszarMills1997,Kahn2010}.
  The odd anharmonicity of the asymmetric stretch is forbidden by hydrogen-exchange symmetry, and the symmetric stretch mode exhibits a larger odd anharmonicity than the bend mode.
  Though both the stretch and bend of the O--H bonds are steeper under compression and softer toward dissociation,
  the bend mode only softens as it approaches linearity, a saddle point $11\,127\,\mathrm{cm}^{-1}$ above equilibrium~\cite{Tarczay1999}.
  Since the saddle point is nearly twenty times the sampled energy window, the bend mode is therefore nearly harmonic at thermal amplitudes.
  In the scans performed, the averages of the residual components are $\langle|\Delta E_2|\rangle = 0.79$ and $1.31$~meV and $\langle|\Delta E_3|\rangle = 0.021$ and $0.055$~meV for bend and stretch respectively, anticipating the ordering and magnitudes of every learning floor in Table~\ref{tab:floors}.

  \begin{table}
    \centering
    \caption{Computed asymptotic error floors for the three water normal-mode scans.
      Each floor is an average over 500 random train/test splits with
      100-point test sets.
      $\epsilon_\infty^{\rm HO}$ is the baseline set by the harmonic-oscillator
      reference labels; $\epsilon_\infty^{\rm Aug_2}$ and $\epsilon_\infty^{\rm Aug_3}$
      are the floors under symmetric and corrected augmentation respectively.}
    \label{tab:floors}
    \begin{tabular*}{\columnwidth}{@{\extracolsep{\fill}}lccc}
      \toprule
      Mode & $\epsilon_\infty^{\rm HO}$ & $\epsilon_\infty^{\rm Aug_2}$ & $\epsilon_\infty^{\rm Aug_3}$ \\
           & (meV) & (meV) & (meV) \\
      \midrule
      Bend        & 0.792 & 0.785 & 0.021 \\
      Sym.\ str.  & 1.321 & 1.324 & 0.056 \\
      Asym.\ str. & 0.055 & ${<}0.001$ & 0.054 \\
      \bottomrule
    \end{tabular*}
  \end{table}

  As one would expect, across all three modes, the qualitative effect of the symmetry is a constant lowering of the log-log offset in the  power-law regime.
  Once training set sizes are sufficiently large that the generalization error approaches the error floor set by the augmentation scheme (Table~\ref{tab:floors}), the default learning takes over.
  This observation is in agreement with Eq.~\eqref{eq:lc_discrete_shift}.

  The learning curves for the two augmented models provide the same performance gain at small $N$, diverging only once they approach their respective convergence floors.
  This illustrates that, in the data-starved regime, higher-order anharmonic errors are unresolved by the model, and that even a second-order harmonic estimate of the mirror label can yield considerable gains.

  For the bending mode~[See Fig.~\ref{fig:water1d}(a)], $\Delta E$ is dominated by its cubic component with a smaller but  nonzero quartic contribution.
  Per Eq.~\eqref{eq:sym_floor}, the $\mathrm{Aug}_2$ floor is set by the odd
  anharmonic component; correspondingly, $\epsilon_\infty^{\mathrm{sym}}$ nearly
  coincides with the HO error (both $\approx 0.79$~meV), since both are dominated by the same cubic
  term in $\Delta E$.
  $\mathrm{Aug}_3$ suppresses the odd contribution, and its floor is instead
  governed by the even component per Eq.~\eqref{eq:taylor_floor}, giving
  $\epsilon_\infty^{\mathrm{corr}}\approx 0.02$~meV.

  The symmetric stretching mode [See Fig.~\ref{fig:water1d}(b)] is dominated by odd-order anharmonicity due to the greater energetic cost of pushing the H atoms closer to the O atom.
  The $\mathrm{Aug}_2$ floor accordingly nearly coincides with the
  HO baseline (both $\approx 1.32$~meV), as both are governed by the same cubic term in $\Delta E$.
  $\mathrm{Aug}_3$ eliminates the leading cubic label error, and the learning curve
  converges to a floor more than an order of magnitude smaller ($0.056$~meV),
  consistent with the subdominant quartic anharmonicity predicted by Eq.~\eqref{eq:taylor_floor}.
  This floor is finite because, while the Hessian correction cancels the cubic error, the O--H stretch carries non-zero quartic anharmonicity.

  The asymmetric stretching mode [See Fig.~\ref{fig:water1d}(c)] is exactly even, since $+q$ and $-q$ elongate opposite O--H bonds and so differ only by permutation of the two H atoms.
  $\mathrm{Aug}_2$ therefore assigns exact labels and its learning curve has no floor, while $\mathrm{Aug}_3$ perturbs an already exact label and incurs the harmonic-baseline error set by the leading quartic term.
  This evenness is permutational invariance rather than a label symmetry, and is already captured by modern atomic representations.

  \subsubsection{Full 3D sampling}

  We apply the augmentation to the full three-dimensional PES of water, sampled
  thermally at $T=300$~K as in standard PES-fitting
  protocols~\cite{ANI_IsayevRoitberg2017, Gastegger2017IR}, under the
  approximate inversion $\mathbf{Q}\mapsto-\mathbf{Q}$ of all three normal
  modes---exact only for the asymmetric stretch, approximate for the bend and
  symmetric stretch.

  \begin{figure}
    \centering
    \includegraphics[width=\linewidth]{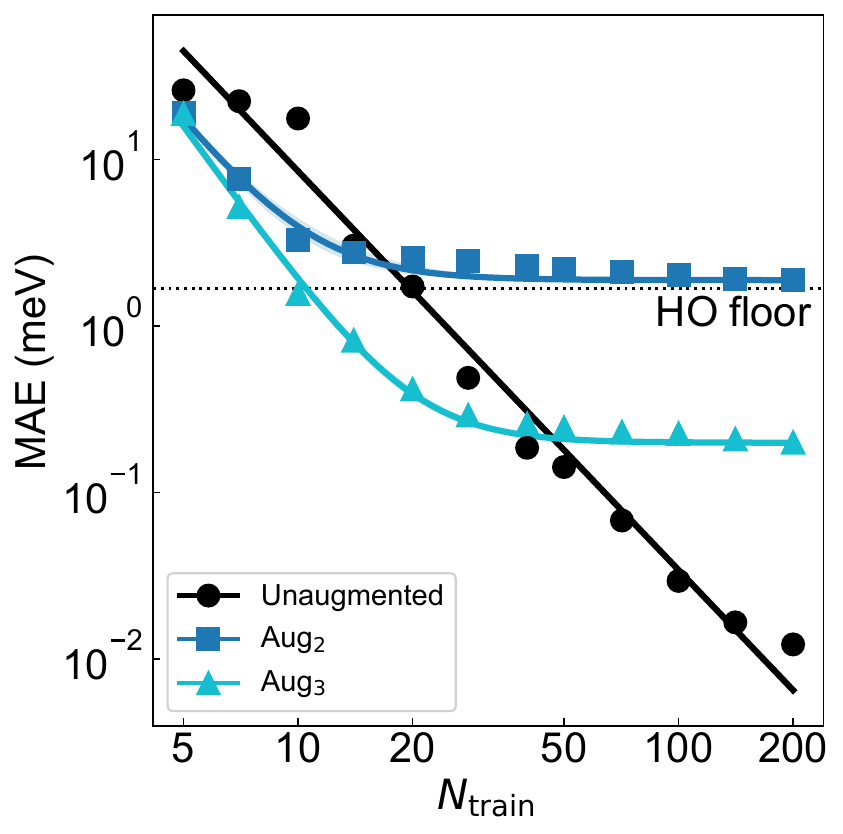}
    \caption{
      Energy learning curves for the three-dimensional water PES for unaugmented, $\mathrm{Aug}_2$,
      and $\mathrm{Aug}_3$ models, evaluated on a $1000$-point test set. The
      unaugmented curve is fit as a power law $\epsilon(N)=A N^{-\beta}$ and the augmented
      curves with the saturating form $\epsilon(N)=\epsilon_\infty + A N^{-\beta}$, where
      $\epsilon_\infty$ is fixed at the MAE of the largest training set. The dotted line
      denotes the harmonic-oscillator baseline ($1.68$~meV).
    }
    \label{fig:water3d}
  \end{figure}

  The learning curves (Fig.~\ref{fig:water3d}) reproduce the floors of
  Sec.~\ref{sec:floors}. $\mathrm{Aug}_2$
  saturates just above the harmonic-oscillator baseline, and $\mathrm{Aug}_3$
  outperforms the unaugmented model by factors ranging between $1.82$ and $1.91$ before
  approaching the floor, close to the twofold gain predicted for reflection symmetry.
  As expected, the
  unaugmented curve overtakes $\mathrm{Aug}_3$ near
  $N \approx 40$.

  Generally, a multidimensional normal mode PES will have a
  mixture of exactly and approximately symmetric axes, so we now discuss the
  usefulness of $\mathrm{Aug}_3$ in this setting. As shown for the asymmetric
  stretch in one dimension, $\mathrm{Aug}_3$ biases a target that is already
  exactly symmetric. However, over the full surface that includes contributions
  from all modes and their couplings, $\mathrm{Aug}_3$ is still advantageous;
  the dominant cubic terms of the bend and symmetric stretch modes are captured
  at the cost of  quartic-order errors in the asymmetric stretch mode terms.
  The $\mathrm{Aug}_3$ floor is then the total
  quartic anharmonicity of every coordinate, including coupling. A quartic
  polynomial fit
  of the
  surface (accurate to $0.02$~meV) supports this, as it is calculated to have
  an even component of $0.16$~meV, close to
  the measured $\mathrm{Aug}_3$ floor of $0.20$~meV.

  Models of the convex well near equilibrium underpin perturbative approaches to
  molecular spectroscopy such as second-order vibrational perturbation theory~\cite{vpt2}.
  Our augmentation supplies
  such a model at reduced cost, with an error floor two
  orders of magnitude below chemical accuracy for water using $\approx 20$
  reference calculations. Energy accuracy is necessary
  but not sufficient for a faithful vibrational spectrum, since fundamental
  frequencies depend crucially on higher-order derivatives of the energy with respect to
  the normal modes. In the next section, we demonstrate the benefit of augmentation
  with approximately symmetric labels for predicting the infrared (IR) spectrum of a single
  water molecule.

  \subsubsection{Anharmonic infrared spectrum}
  \label{sec:ir}

  The IR fundamentals provide a derivative-level test of the learned PES. In VPT2,
  the peak positions are controlled by the local cubic and quartic force field, which
  we extract by analytic differentiation of each KRR model. Since the corrected
  $\mathrm{Aug}_3$ label already uses the Hessian [Eq.~\eqref{eq:taylor_mirror_label}],
  we learn the anharmonic residual $\Delta E = E-E_\mathrm{HO}$ directly with
  a $\Delta$-learning model~\cite{deltaML2015}. This approach fixes the harmonic
  frequencies exactly and isolates the effect of the learned
  anharmonicity on peak displacements.

  For each trained model, we pass the harmonic plus extracted residual quartic surface to
  pyVPT2.
  Table~\ref{tab:ir} tabulates the harmonic, cubic, and quartic contributions to the
  reference VPT2 fundamentals, as well as the peak positions obtained from the
  representative KRR spectra at $N=128$.
  The cubic and quartic contributions are opposite
  in sign, with the cubic red-shifting the stretches and the quartic blue-shifting them,
  and vice versa for the bend.
  This decomposition also explains the convergence of each
  augmentation scheme.
  The $\mathrm{Aug}_2$ model is even under normal-coordinate inversion and therefore
  cannot recover the odd cubic force field.
  In principle it can represent even
  quartic terms, but in practice the cubic label noise results in strongly regularized
  learned quartic derivatives, leaving the spectrum close to the harmonic baseline.
  $\mathrm{Aug}_3$ rapidly resolves the cubic shifts
  but converges to a cubic-only spectrum due to the fourth-order error of the correction.
  The unaugmented model has no symmetry-imposed
  floor and should converge to the full VPT2 reference once sufficient
  data are available.

  \begin{table*}
    \centering
    \caption{Cubic and quartic contributions to the water VPT2 fundamentals, together with
      the learned fundamentals ($\mathrm{cm}^{-1}$) at training set size $N=128$. The harmonic VPT2 baseline plus the
      cubic and quartic increments sum to the DFT/VPT2 reference. The final three
      columns give the peak positions from the same representative spectra shown in
      Fig.~\ref{fig:ir_spectrum} at $N=128$.}
    \label{tab:ir}
    \begin{tabular*}{\textwidth}{@{\extracolsep{\fill}}lccccccc}
      \toprule
      & \multicolumn{4}{c}{VPT2 decomposition} & \multicolumn{3}{c}{Representative model ($N=128$)} \\
      \cmidrule(lr){2-5}\cmidrule(lr){6-8}
      Mode & Harmonic & $+$cubic & $+$quartic & Reference & $\mathrm{Aug}_2$ & $\mathrm{Aug}_3$ & Unaug. \\
      \midrule
      Bend        & 1654.3 & $+26.4$  & $-109.0$ & 1571.7 & 1654.2 & 1685.4 & 1580.9 \\
      Sym.\ str.  & 3894.3 & $-320.1$ & $+150.1$ & 3724.3 & 3894.0 & 3503.0 & 3732.6 \\
      Asym.\ str. & 4012.9 & $-340.9$ & $+133.5$ & 3805.4 & 4012.7 & 3625.0 & 3824.9 \\
      \bottomrule
    \end{tabular*}
  \end{table*}

  \begin{figure}
    \centering
    \includegraphics[width=\linewidth]{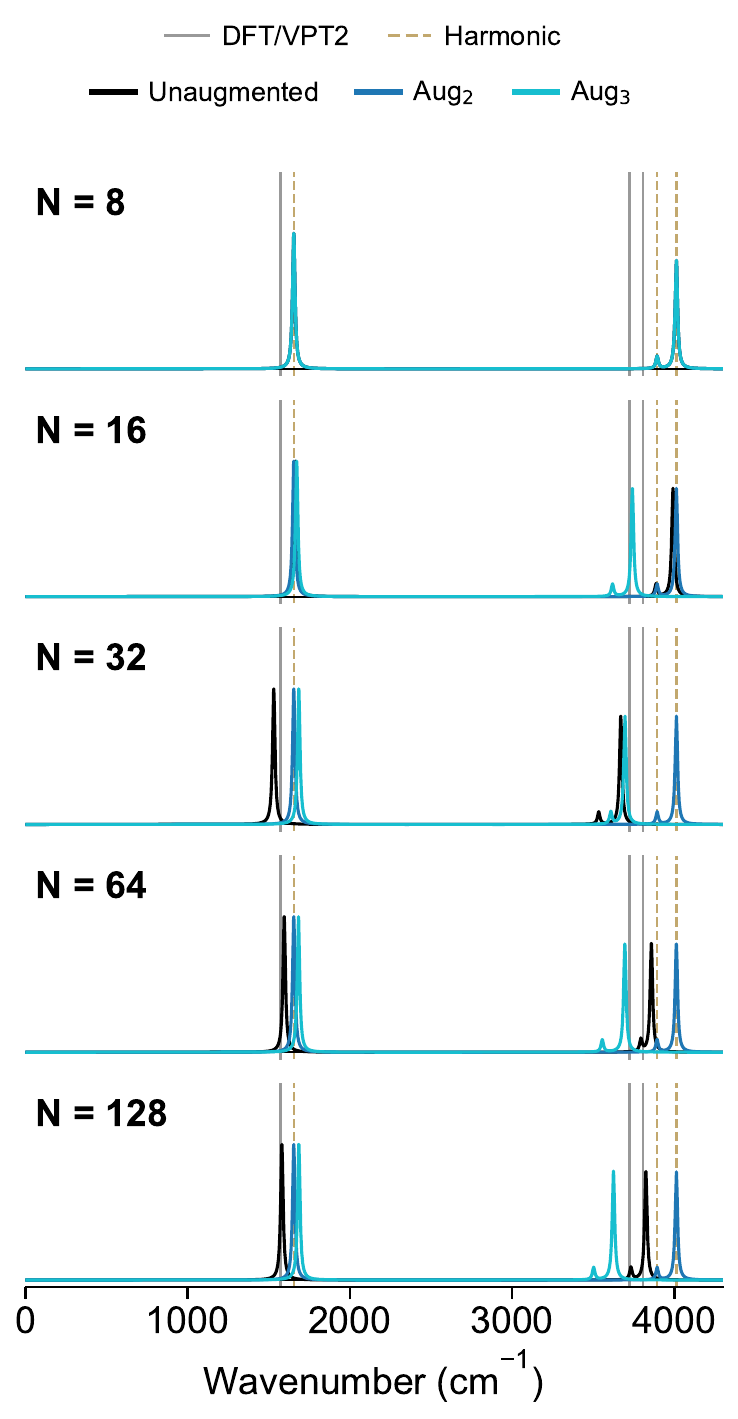}
    \caption{
      Predicted anharmonic infrared spectra of water from the $\Delta$-learning model
      as a function of training-set size $N$ (rows $N=8$--$128$), under unaugmented
      (black), $\mathrm{Aug}_2$ (blue), and $\mathrm{Aug}_3$ (cyan) labeling. Grey
      solid lines mark the DFT/VPT2 reference fundamentals and gold dashed lines the
      harmonic VPT2 fundamentals. Spectra are broadened with Lorentzians
      ($\gamma=18\,\mathrm{cm}^{-1}$) and each is normalized to its maximum.
    }
    \label{fig:ir_spectrum}
  \end{figure}

  We demonstrate how the force constants are learned by tracking the progression of IR
  spectra obtained from representative KRR models (Fig.~\ref{fig:ir_spectrum}).
  At $N=8$, all schemes collapse to the harmonic baseline, illustrating that the higher-order
  derivatives at equilibrium are not yet resolved.
  $\mathrm{Aug}_2$ never deviates from harmonic even though it has the correct symmetry
  to learn the quartic contributions; the odd
  label noise from the cubic terms results in a highly regularized model that suppresses
  the quartics at every $N$.
  $\mathrm{Aug}_3$ recovers the leading cubic shifts already by $N=16$: the stretch
  peaks move strongly to the red, while the bend moves to the blue, matching the
  signs of the cubic contributions in Table~\ref{tab:ir}.
  As expected, its large-$N$ spectrum is limited by the omitted
  quartic contribution.
  The unaugmented model also recovers first the larger cubic correction, visible by $N=32$,
  and only at larger training sizes does it learn the quartic correction to converge to the
  full VPT2 reference.

  Our results suggest that approximate label symmetries can be useful even for
  derivative-level observables.
  The advantage is not that the augmented model is
  asymptotically exact, but rather that the leading symmetry-breaking terms can be
  accessed with less reference data.
  When the
  observable is dominated by that term, as the stretch fundamentals are dominated
  by the cubic force field here, the augmented labels recover the qualitative
  anharmonic physics with half the single-point calculations.
  Approximate
  augmentation is therefore most useful as a low-data, physically interpretable
  bias for resolving the leading correction to a symmetric reference, while the
  remaining deviations expose which higher-order terms still need to be learned.

  \section{Conclusion}

  We have studied the use of label symmetry as an inductive bias for supervised learning in both exact and approximate settings.
  Exact symmetries have been illustrated for the hydrogen atom's electronic orbital densities which lead to expected improvements in learning-curve behavior.
  Continuous rotational invariance of the $s$-orbital
  density increases the convergence rate (steeper slope) by reducing the effective input dimensionality from three to one, while the discrete reflection symmetries of the
  $p_z$ and $d_{xz}$ densities shifts the generalization error's pre-factor without altering  the slope, with each fold contributing approximately a factor of two to the effective training set size.

  For approximate label symmetries, the gain is bounded by the symmetry-breaking  terms.
  In the water potential energy surface near equilibrium,
  normal-mode reflection symmetry is broken by anharmonicity.
  Symmetric augmentation therefore improves the low-data regime but
  converges to a floor set by the leading cubic anharmonicity. Corrected augmentation
  includes a Hessian-based correction that
  cancels the leading cubic error and shifts the error floor to quartic order.
  In both the one-dimensional scans and the full three-dimensional surface, the
  learning curves show the expected twofold gain for reflection symmetry; the
  advantage window is set by the error floor of the chosen
  augmentation scheme.

  Downstream task observables, such as vibrational frequencies also benefit
  from approximate reflection symmetry. Corrected augmentation recovers the
  dominant cubic shifts with half the single-point calculations, and its
  error is the omitted quartic contribution. Consequently, approximate label
  symmetry is most useful as a low-data bias for resolving the leading correction
  to a symmetric reference, rather than as an asymptotically exact constraint.

  In conclusion, exploiting approximate label symmetries holds promise in data scarce regimes and whenever the target is organized around a symmetric reference points, such as a local minimum. Under such conditions, augmentation can recover the dominant
  correction before an unconstrained model has enough data, while the remaining
  floor identifies the higher-order physics left out.
  Future work will explore whether approximate reflection symmetry gives similar
  data-efficiency gains in other materials settings, such as anharmonic lattice
  dynamics, and whether other weakly broken label symmetries can be exploited
  in the same way.

  \begin{acknowledgments}
  We acknowledge the support of the Natural Sciences and Engineering Research Council of Canada (NSERC), [funding reference number RGPIN-2023-04853]. Cette recherche a été financée par le Conseil de recherches en sciences naturelles et en génie du Canada (CRSNG), [numéro de référence RGPIN-2023-04853].
This research was undertaken thanks in part to funding provided to the University of Toronto's Acceleration Consortium from the Canada First Research Excellence Fund,
grant number: CFREF-2022-00042.
O.A.v.L. has received support as the Ed Clark Chair of Advanced Materials and as a Canada CIFAR AI Chair.
  \end{acknowledgments}

  \bibliography{main}

\clearpage
\appendix
\renewcommand{\thefigure}{S\arabic{figure}}
\setcounter{figure}{0}
\renewcommand{\thetable}{S\arabic{table}}
\setcounter{table}{0}
\renewcommand{\theequation}{S\arabic{equation}}
\setcounter{equation}{0}

\section{Hydrogen $2s$ and $3s$ learning curves}

\begin{figure*}
  \centering
  \includegraphics[width=0.9\linewidth]{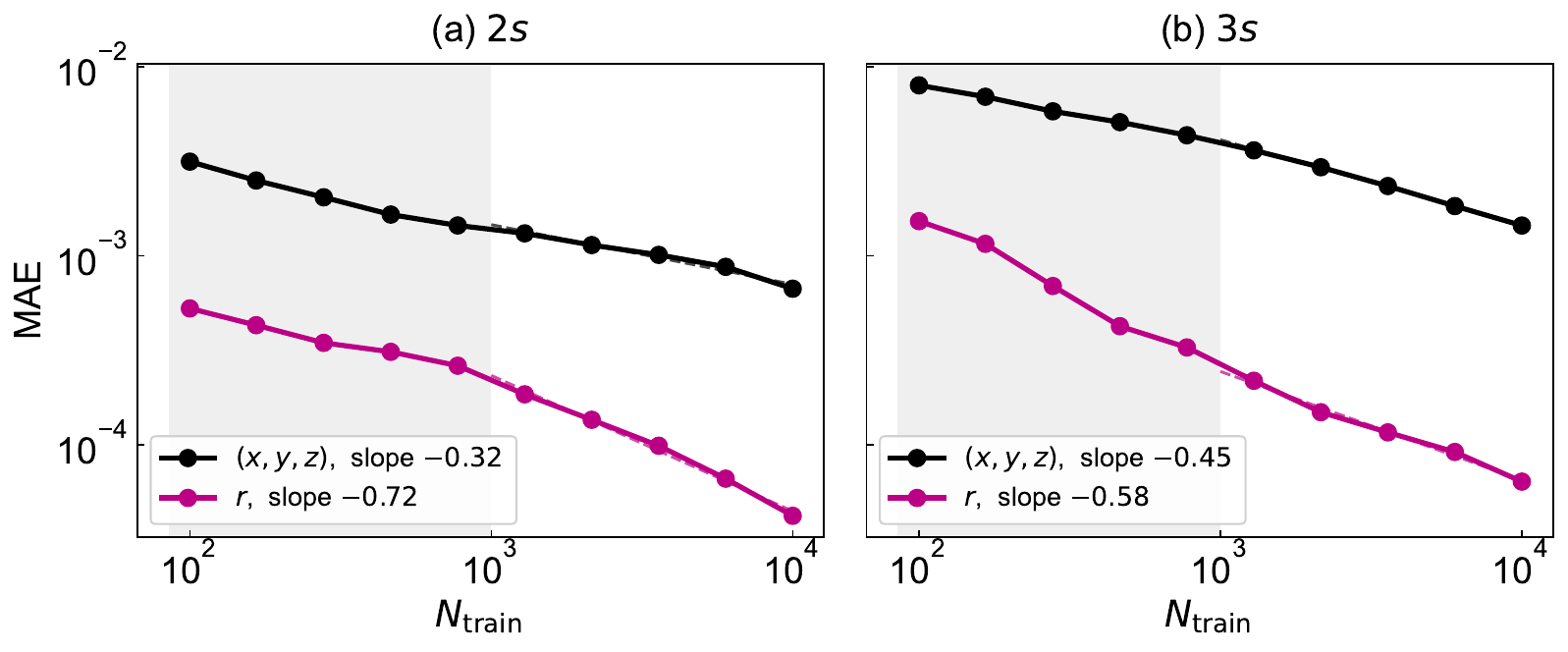}
  \caption{Learning curves for the (a) $2s$ and (b) $3s$ hydrogen orbital densities under direct training on Cartesian coordinates $(x,y,z)$ (black) and radial training on the rotation-invariant coordinate $r$ (magenta). Power-law slopes, given in each legend, are fitted over the approximately linear regime $N_\mathrm{train}\geq 1000$, with the shaded small-$N$ region excluded from the fit. The dashed lines show the fitted power laws. These mirror the $1s$ result shown in Fig.~2 of the main text.}
  \label{fig:si_sorbitals}
\end{figure*}

Figure~\ref{fig:si_sorbitals} shows the learning curves for the $2s$ and $3s$ hydrogen orbital densities, which are summarized in the main text but omitted from Fig.~2 for compactness. As with the $1s$ density, the model trained on the rotation-invariant radial coordinate $r$ converges faster than the direct model trained on Cartesian coordinates $(x,y,z)$, consistent with the effective dimensionality reduction afforded by the continuous $O(3)$ label symmetry. The fitted log-log power-law slopes are given in each legend. They are obtained over the asymptotic window $N_\mathrm{train}\geq 1000$ where the curves are approximately linear, with the small-$N$ region shaded and excluded from the fit. The radial models are steeper for both orbitals, consistent with the $1s$ result in the main text. At $N_\mathrm{train}=10\,000$ the radial input yields a data-efficiency gain of $15.9\times$ for the $2s$ density and $22.7\times$ for the $3s$ density.

\section{Equivalence of symmetry-enforcement strategies}

\begin{figure}[h]
  \centering
  \includegraphics[width=0.7\linewidth]{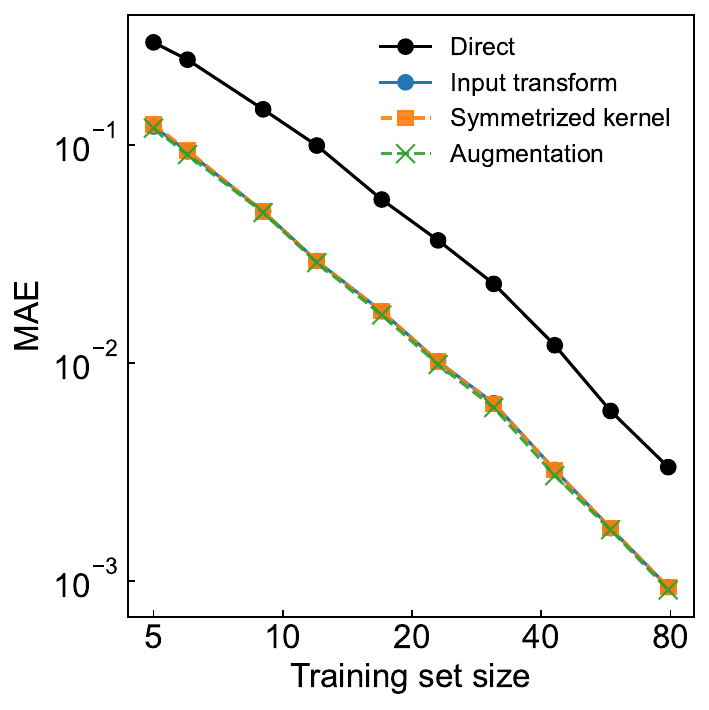}
  \caption{Learning curves for the even target $f(x)=x^2$ under the reflection
  symmetry $x\mapsto-x$, comparing the unconstrained model (direct) with three
  strategies for enforcing the symmetry, namely input transformation, a symmetrized
  kernel, and data augmentation. The three symmetry-adapted curves coincide and are
  offset from the direct curve by a constant factor, as predicted for a discrete
  symmetry. Mean absolute error (MAE) is computed on a held-out test set.}
  \label{fig:parabola_methods}
\end{figure}

For an exact symmetry, data augmentation, input transformation, and kernel
symmetrization enforce the same invariance and yield the same kernel ridge
regression predictor. Figure~\ref{fig:parabola_methods} confirms this for a
one-dimensional even target $f(x)=x^2$, whose reflection symmetry $x\mapsto-x$ is
exact. The learning curves for the three strategies coincide to within line width
across the full range of training set sizes, and all lie below the curve of the
unconstrained (direct) model by the constant log-log offset expected for a
discrete symmetry. Because the strategies are equivalent, the choice among them in
the main text is made purely for convenience.

\section{Force-constant recovery under symmetric augmentation}

\begin{figure*}
  \centering
  \includegraphics[width=\linewidth]{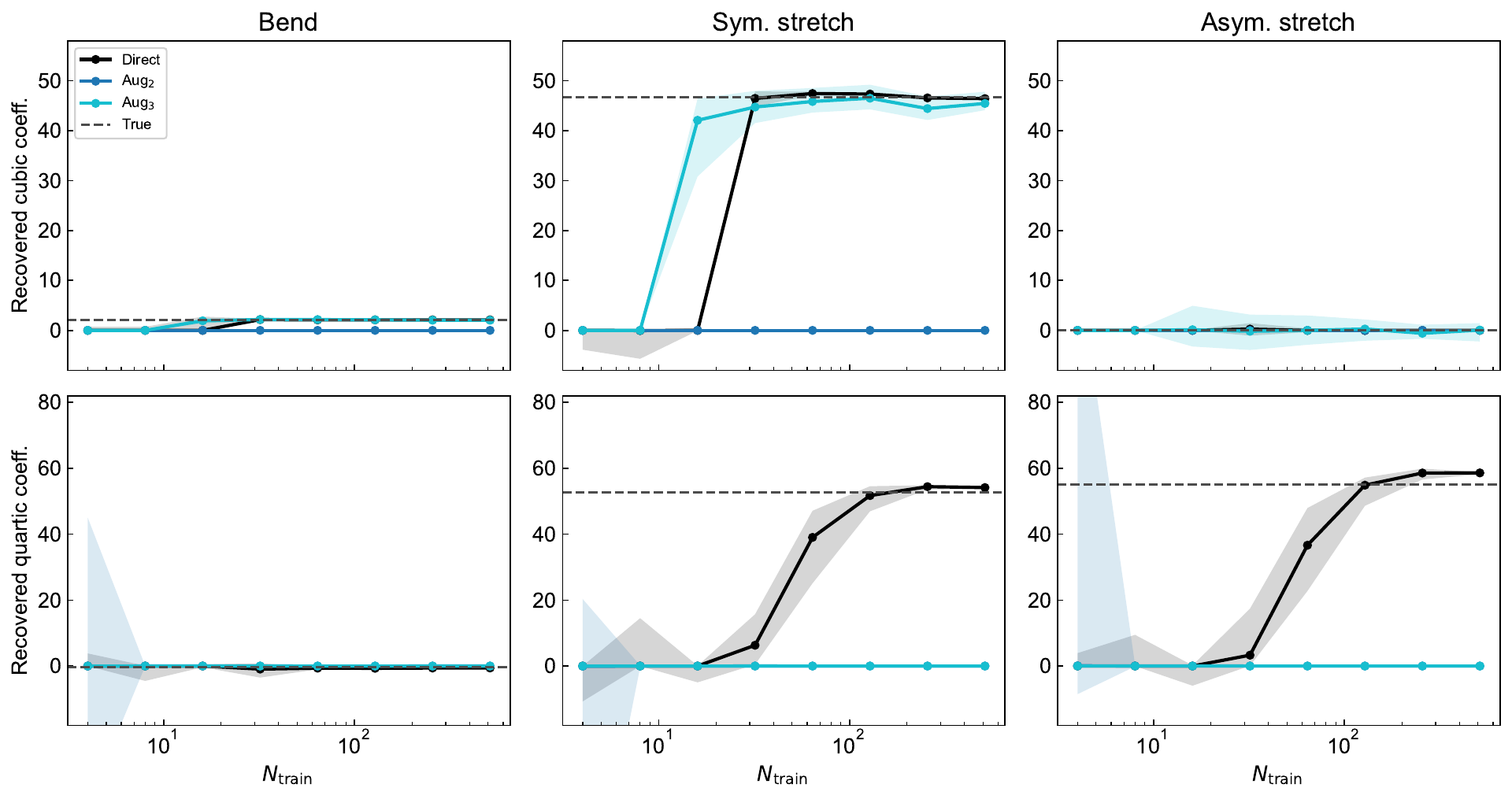}
  \caption{Recovered cubic (top row) and quartic (bottom row) Taylor
  coefficients of the $\Delta$-learning energy model for the bend, symmetric
  stretch, and asymmetric stretch normal modes (columns), as a function of
  training-set size $N_\mathrm{train}$, under direct (black), $\mathrm{Aug}_2$
  (blue), and $\mathrm{Aug}_3$ (cyan) labeling. Lines and shaded bands are the
  median and interquartile range over training selections. Dashed lines mark the
  true values from a quartic fit to the DFT residual. The direct model recovers
  both force constants, $\mathrm{Aug}_3$ recovers the cubic and zeros the quartic
  by construction, and $\mathrm{Aug}_2$ recovers neither.}
  \label{fig:recovered_fc}
\end{figure*}

Figure~\ref{fig:recovered_fc} reports the cubic and quartic force constants
recovered by the $\Delta$-learning energy models, obtained by analytic
differentiation of the trained model at the equilibrium geometry and compared
with the true values from a quartic fit to the DFT residual. The direct model
recovers both force constants once $N_\mathrm{train}\gtrsim 100$. $\mathrm{Aug}_3$
recovers the cubic earliest, by $N_\mathrm{train}\approx 16$, and zeros the
quartic by construction. $\mathrm{Aug}_2$ recovers neither force constant,
removing the cubic exactly by symmetrization and failing to resolve the quartic
because the discarded odd anharmonicity acts as a label noise on the even fit.
The figure substantiates the parity analysis of the infrared spectrum in the
main text.

\end{document}